\documentclass[%
 reprint,
superscriptaddress,
 amsmath,amssymb,
 aps,
]{revtex4-2}
\usepackage{graphicx}
\usepackage{dcolumn}
\usepackage{bm}
\usepackage{relsize}
\usepackage{url} 
\usepackage[utf8]{inputenc}
\usepackage{comment}
\usepackage{amsmath}
\usepackage{amssymb}
\usepackage{bm}
\newcommand{\bt}[1]{\mathbf{#1}}

\usepackage{xcolor}
\usepackage[pdftex, pdftitle={Article}, pdfauthor={Author},colorlinks,citecolor=blue,linkcolor=blue,urlcolor=blue]{hyperref}
\usepackage{cleveref}

\begin{document}

\title{Topological Superconductivity in Twisted Flakes of Nodal Superconductors}

\author{Kevin P. Lucht}
\affiliation{%
 Department of Physics and Astronomy, Center for Materials Theory,
Rutgers University, Piscataway, NJ 08854, USA
}%
 \author{J. H. Pixley}
\affiliation{%
 Department of Physics and Astronomy, Center for Materials Theory,
Rutgers University, Piscataway, NJ 08854, USA
}%
\affiliation{
Center for Computational Quantum Physics, Flatiron Institute, 162 5th Avenue, New York, NY 10010
}%
\author{Pavel A. Volkov}%
\affiliation{%
Department of Physics, University of Connecticut, Storrs, Connecticut 06269, USA
}%
\affiliation{%
 Department of Physics and Astronomy, Center for Materials Theory,
Rutgers University, Piscataway, NJ 08854, USA
}%
\affiliation{%
Department of Physics, Harvard University, Cambridge, Massachusetts 02138, USA
}%

\date{\today}

\begin{abstract}
Twisted bilayers of nodal superconductors have been recently demonstrated to be a potential platform to realize two-dimensional topological superconductivity. Here we study the topological properties of twisted finite-thickness flakes of nodal superconductors under applied current, focusing on the case of a $N$-layer flake with a single twisted top layer. At low current bias and small twist angles, the average nodal topological gap is reduced with flake thickness as  $\sim\mathcal{O}(\frac{1}{N})$, but the Chern number grows $\sim \mathcal{O}(N)$. As a result, we find the thermal Hall coefficient to be independent of $N$ at temperatures larger than the nodal gap. At larger twist angles, we demonstrate that the nodal gap in the density of states of the top layer is only weakly suppressed, allowing its detection in scanning tunneling microscopy experiments. These conclusions are demonstrated numerically in an atomic-scale tight-binding model and analytically through the model's continuum limit, finding excellent agreement between the two.
Finally, we show that increasing the bias current leads to a sequence of topological transitions, where the Chern number increases like $\sim\mathcal{O}(N^2)$ beyond the additive effect of stacking $N$ layers. Our results show that twisted superconductor flakes are ``$2.5$-dimensional'' materials, allowing to realize new electronic properties due to synergy between two-dimensional layers extended to a finite thickness in a third dimension.
\end{abstract}

\maketitle


\section{Introduction}
The realization of twisted two-dimensional (2D)  moir\'{e} materials has 
led to several discoveries of novel correlated electronic phases ~\cite{Kim2017,Cao2018}.
The field has advanced from twisted bilayer graphene (TBG) to new discoveries in hBN substrate-aligned TBG~\cite{balents2020} and twisted graphene multilayers~\cite{Chen2019,Shen2020,Zhang2022,Burg2022,Shen2023}, magnetic moir\'{e} structures~\cite{Tiancheng2021}, and  twisted or stacked transition metal dichalgonides (TMDs)~\cite{Wang2020,Regan2020}. In particular, TMDs have been
instrumental in observing anomalous topological phases such as a quantum anomalous Hall effect likely due to orbital ferromagnetism in heterostructures~\cite{Li2021}, and (fractional) Chern insulators in twisted bilayers~\cite{Cai2023,Zeng2023,Xu2023}. 
Recently, the twist paradigm has been extended to
2D nodal superconducting materials, leading to proposals that realize chiral topological superconductivity~\cite{Can2021,Volkov2022} and spontaneous time-reversal symmetry breaking~\cite{Can2021,Zhao2021,Volkov2022_2,Volkov2021,Lu2022,Song2022}.

The unusual properties of twisted bilayers have motivated proposals for multilayer structures~\cite{Khalaf2019,Tummuru2022}, going all the way to three dimensions (3D)~\cite{sarma3d} to realize new electronic states, including topological ones.
Theoretically, layered chiral topological systems can realize new types of behavior including edge sheaths for integer quantum Hall multilayers~\cite{chalker1995,balents1996}. Moreover, their extension to interacting systems have been predicted to form new types of topological ordered phases in 3D ~\cite{balents1996,Levin2009,sondhi2000,Gooth_2023}.

Another approach to extend the possible phenomena of 2D materials is to consider twisted flakes of finite thickness. Remarkably, recent experiments have demonstrated that even a single layer of graphene twisted on top of a finite-thickness graphite flake can modify the electronic structure of the whole flake~\cite{waters2023mixed,mullan2023mixing}. Single layer twists represent a simple but novel route to realize effective ``2.5D'' heterostructures that truly lie in between two and three dimensional behaviour.

For twisted superconducting platforms~\cite{Zhao2021,yejin2023,martini_2023}, there is also a practical motivation to study multilayers. While nodal superconductivity in individual bi- and mono-layers has been achieved~\cite{frank2019,Yu2019}, realizing twisted bilayers remains a challenge. Instead, experiments focused on the interfaces between finite-thickness flakes~\cite{Zhao2021,yejin2023,martini_2023}. Such a setup differs from the proposed twisted multilayer structures~\cite{Khalaf2019,Tummuru2022}, where the twist angle between each pair of layers has to be controlled. 

\begin{figure*}[t]
  \includegraphics[width=.95\textwidth]{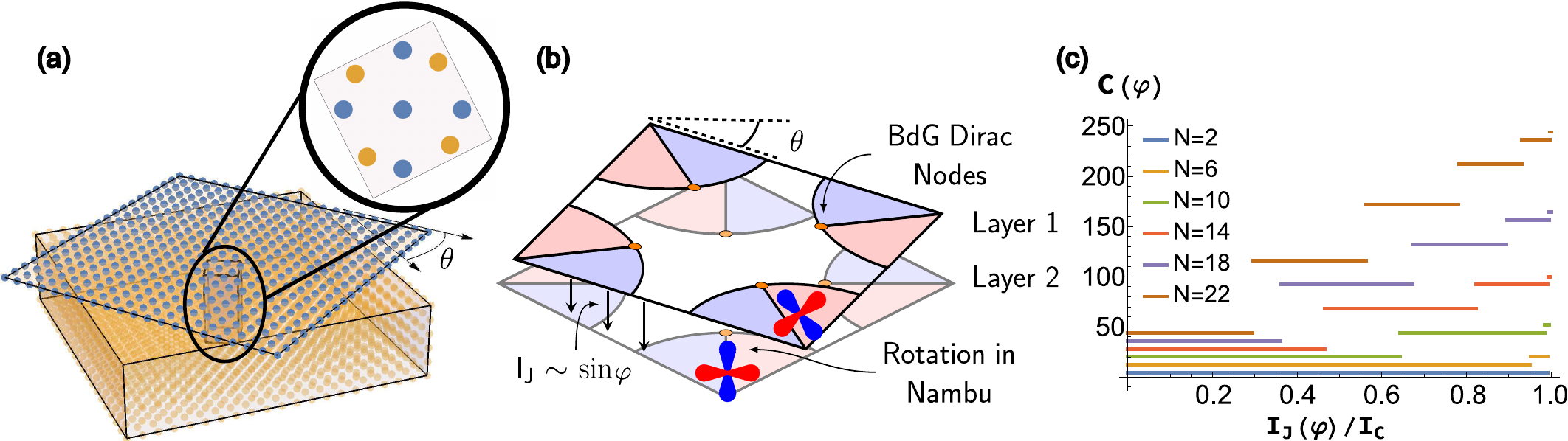}
  \caption{(a) Illustration of the $N$-layer flake for $\theta_{1,2}\approx 53^\circ$. Inset: single moir\'{e} unit cell. 
  (b) Momentum-space illustration for $N=2$. Shown are the Fermi surfaces of each layer with a $d_{x^2+y^2}$ singlet order parameter, with its sign marked by color and quasiparticle nodes highlighted in orange. Combination of momentum-space rotation and interlayer supercurrent $I_J$ produces a Nambu-space rotation of the order parameter to a $d+id$ topological one.
  (c) Phase diagram of $N$-layered systems for varying current and twist angles $\theta < \theta_{MA} $. Shown are the predicted net Chern numbers ($C(\varphi)$) as a function of normalized Josephson current ($\frac{I_J(\varphi)}{I_c}$). For $N \leq 5$, $C$ grows linearly with $N$ and is independent of $I_J$, but for $N \geq 6$, topological phase transitions can occur increasing $C$ to values up to $\mathcal{O}(N^2)$ for $N\gg1$.
  }
\label{fig:2_layer_cartoon_and_Chern}
\end{figure*}

In this work, we consider the possibility to induce topological superconducting states in two twisted multilayer flakes of nodal superconductors. Focusing on the case of a single twisted layer on top of a $N$-layer flake (Fig. \ref{fig:2_layer_cartoon_and_Chern} (a) and (b)), we demonstrate that the twisted interface induces topological superconductivity in the whole flake with an interlayer supercurrent. While the topological gap is reduced in flakes as $\mathcal{O}(1/N)$, we identify experimentally accessible observables (density of states and thermal Hall effect) that are thickness-independent. Unexpectedly, we find that increasing the current flowing through the stack leads to a sequence of topological transitions, reconstructing the quasiparticle spectrum of the whole flake. As a result, the Chern number of the system can be increased up to $\mathcal{O}(N^2)$ (Fig. \ref{fig:2_layer_cartoon_and_Chern} (c)), beyond the expectation from $N$ layers of topological superconductors, opening the path to intermediate, "2.5D" topological superconductivity.

The paper is organized as follows. We present two models of the system in Sec. \ref{sec:models}: a microscopic model of a moir\'{e} superlattice and a low energy continuum model of a single valley of $N$ Dirac nodes. In Sec. \ref{sec:cont}, we first explore small and large twist angle limits to understand the evolution of topological superconductivity with $N$. Sec. \ref{sec:exp} then introduces the experimental observables: density of states and thermal Hall effect, and highlights the predictions of the model which may be probed in experiments. We end with Sec. \ref{sec:conclusion} to summarize the results and comment on the emergence of 2.5D phenomena in the twisted superconducting flakes.

\section{Models}
\label{sec:models}
We utilize two models to approach the $N$ layered system: a real-space tight-binding model of the twisted multilayer~\cite{Can2021} and an approximate momentum-space model around the Dirac gap nodes~\cite{Volkov2022}. In both cases, we assume that interlayer tunneling is weak compared to the superconducting gap maximum, allowing us to neglect the corrections to the self-consistency equation due to tunneling~\cite{Volkov2022_2} and treat the order parameter as fixed. We use the first model to exactly include  the effects of moir\'{e} reconstruction of the bands and to study the topological edge modes, but limit its use to large commensurate twist angles $\theta>\theta_{MA}$ where the moir\'{e} unit cell is not too large, and moderate values of $N$. The continuum model, on the other hand, allows to study effects of arbitrary $N$ and twist angles analytically. This allows us to compare the results of two models at large twist angles where we find good agreement in the overlapping regimes of applicability.

\subsection{Lattice Model}
\label{subsec:lattice} 

To model the system at the microscopic lattice level, we use a tight-binding approach. In absence of interlayer tunneling, the Hamiltonian takes the form:
\begin{equation}
\begin{gathered}
H_{\mathrm{intra}}  =  -t \sum_{\langle i,j \rangle, n, \sigma} ( c_{in\sigma}^\dagger c_{jn\sigma} +  \mathrm{h.c.} ) -\mu  \sum_{i,n,\sigma} n_{in\sigma}
\\
+
\sum_{\langle i,j \rangle, l} ( \Delta_{ij}e^{i\varphi_n} c_{in\uparrow}^\dagger c^\dagger_{jn\downarrow} + \mathrm{h.c.}),
\label{eqn:OP_NN}
\end{gathered}
\end{equation}
where  $n, \sigma$ and $i,j$ are the layer, spin and site indices, respectively, and $\langle i,j \rangle$ is a sum over nearest neighboring sites on the square lattice. $t$ and $\mu$ are the single-electron hopping and chemical potential, while $\Delta_{ij}$ is the $d$-wave SC order parameter 
\begin{equation*}
    \Delta_{ij} =  \begin{cases}
\phantom{-} \Delta \quad \textrm{if } \langle ij \rangle = \hat{x} \\
-\Delta \quad \textrm{if } \langle ij \rangle = \hat{y}
\end{cases}.
\end{equation*}
The square lattice sites for $n=1$ are rotated by an angle $\theta$ (see Fig. \ref{fig:2_layer_cartoon_and_Chern} (a)). 
The single-electron hopping between the layers is described by:
\begin{eqnarray}
H_{\mathrm{tun}} =
 -\sum_{ i,j ,\sigma} \sum_{n=1}^{N-1} g_{ij} ( c_{in\sigma}^\dagger c_{j,n+1,\sigma} +\mathrm{h.c.}),
\end{eqnarray}
where $ g_{ij} = g_0 e^{-(r_{ij} - d)/\rho}$, $g_0$ being the overall tunneling magnitude and $r_{ij} = \sqrt{d_{ij}^2 +d^2}$ with $d_{ij}$ the in-plane distance between sites and $d$ is the interlayer distance along the z-axis. We do not take the dependence of tunneling on the twist angle due to orbital structure relevant for cuprates \cite{Song2022}, focusing on the qualitative effects of increasing $N$.
We use parameters similar to previous studies: $\Delta=40$ meV~\cite{Can2021}, $t = -126$ meV and $\mu = 15$ meV ~\cite{Markiewicz2005,Volkov2021}, $\rho =2.11 \text{ \AA}$ and $d = 12 \text{ \AA}$ is the interlayer separation~\cite{Can2021}. We also restrict our numerical analysis to twist angles where the system is strictly periodic: $\theta_{p,q} = 2\arctan\left(\frac{p}{q}\right)$, where $p,q\in \mathbb{N}$. We implement the twist by rotating square lattices of the top and remaining $N-1$ layers in opposite directions by $\theta_{p,q}/2$.

Finally, we consider the application of a supercurrent along the stack, shown to induce topology in bilayers~\cite{Volkov2022,Volkov2022_2} displayed in Fig. \ref{fig:2_layer_cartoon_and_Chern} (b). The current $I_J(\theta)$ between two twisted layers is related to the phases of the order parameter by the Josephson relation ~\cite{Volkov2021}:
\begin{equation}
    I_J(\theta) \approx  I_c(\theta)   \sin(\varphi_1-\varphi_2), 
    \label{eq:josephson}
\end{equation}
\noindent where $I_c(\theta)$ is the critical current which depends on the twist angle between the two layers, and $\varphi_n$ is the phase of the order parameter where $n=1,2$ is the layer index.
In experiments on twisted flakes of BSCCO~\cite{Zhao2021}, the critical current was found to follow the expected $d$-wave form $I_c(\theta) \approx I_c \cos (2\theta), $ consistent with predictions for inhomogeneous tunneling~\cite{Volkov2021}. Therefore, the critical current between the top and next layer is reduced with respect to the rest of the stack.  A constant Josephson current then corresponds  to $\varphi_n = n\varphi$ , for layer $n$ with $n\neq 1$ and $\varphi_1 = 2\varphi- \varphi_c$ for the top layer ($n=1$) where $\varphi_c = \arcsin(\frac{\sin\varphi}{ \cos( 2\theta_{p,q})} )$.

\subsection{Continuum Model}
To make the problem analytically tractable while also accessing small twist angles and arbitrary $N$,
we construct a low energy continuum model of the system depicted in detail for $N=2$ in Fig. \ref{fig:2_layer_cartoon_and_Chern} (b) and generalize it to an $N$ layer flake shown in Fig. \ref{fig:kz_nodes} (a). We focus on the case of singlet superconductors, with extensions to triplet being discussed in Ref. \cite{Volkov2022_2}. Each layer of the system is characterized by a single-particle dispersion $\varepsilon(\bt{k}) \tau_3$ and a superconducting order parameter $\Delta(\bt{k})\tau_1$ where $\tau_i$ represent matrices in Gor'kov-Nambu space.

Near the nodes, where $\varepsilon(\bt{K}_N) = 0$, $\Delta(\bt{K}_N) = 0$, we expand to linear order and define variables \cite{Volkov2022_2},
\begin{equation}
    \varepsilon(\bt{K}) \approx \bt{v}_F \cdot (\bt{K}- \bt{K}_N ) \equiv \xi,
    \label{eq:xi}
\end{equation} 
\noindent and 
\begin{equation}
    \Delta(\bt{K}) \approx \bt{v}_\Delta \cdot (\bt{K} - \bt{K}_N ) \equiv \delta,
    \label{eq:delta}
\end{equation}
where $\bt{v}_F$ and $\bt{v}_\Delta$ are the velocities of the normal dispersion and the order parameter, respectively. Since $\bt{v}_F = v_F \hat{k}_\parallel$ and $\bt{v}_\Delta = v_\Delta \hat{k}_\perp$ where $\hat{k}_\parallel$ is the direction parallel to $\bt{K}_N$ and $\hat{k}_\perp$ the direction perpendicular, $\xi$ and $\delta$ can be treated as an orthogonal coordinate system local to $\bt{K}_N$.
The interlayer tunneling can be approximated by a constant term $t \tau_3$~\cite{Volkov2022_2}. For a twist at the top layer, we also include a contribution due to a momentum shift ${\bt Q}= -\theta[\bt{K}_N\times z]$. We further consider the case of a circularly symmetric dispersion $\varepsilon$; at $\theta\ll\theta_{MA}$ non-circular corrections are negligible, while at large twists their effect does not alter the qualitative behavior in twisted bilayers~\cite{Volkov2022_2}. Therefore, the twist contributes a term $\delta_0\tau_1$ where $\delta_0 = \bt{v}_\Delta \cdot {\bt Q}$ between the top layer and its neighboring layer. We also assume $\theta\ll1$, such that the $\theta$ -dependence of $I_c$, Eq.~(corrections $O(\theta^2)$) can be neglected and a uniform supercurrent along the stack corresponds to $\Delta_n({\bf k})\to \Delta_n({\bf k}) e^{i\varphi n}$ where $n$ is the layer index. The resulting Hamiltonian in the vicinity of a single node within each layer takes the form
\begin{equation}
\begin{gathered}
 H(\bt{k},\varphi) = \sum_{n=1}^{N} \xi(k_\parallel) \Phi^\dagger_{n} \tau_3  \Phi_{n}   - t \sum_{n=2}^N \Phi^\dagger_{n}  \tau_3  \Phi_{n+1} + h.c. 
 \\
\sum_{n=1}^{N} [\delta(k_\perp)+\delta_0 \delta_{n,1}]  \Phi^\dagger_{n} [\cos(\varphi n) \tau_1 - \sin(\varphi n)\tau_2] \Phi_{n}.
    \label{eqn:Hcont_full}
    \end{gathered}
\end{equation}
where $\Phi^\dagger_{n} = (c^\dagger_{\uparrow,n}, c_{\downarrow,n} ) $ are the Nambu spinors for layer $n$, $t$ is the tunneling strength, and $\delta_{n,1}$ is the Kronecker delta acting in layer space.  Viewing the layer indexing as sites, this model can be viewed as a quasi-one-dimensional chain with $\xi$ and $\delta$ acting as parameters entering from the perpendicular plane.

\section{Results}
\label{sec:cont}

\subsection{Small twist angles}
\label{subsec:small_angle}

\begin{figure}[h!]
  \includegraphics[width=.4\textwidth]{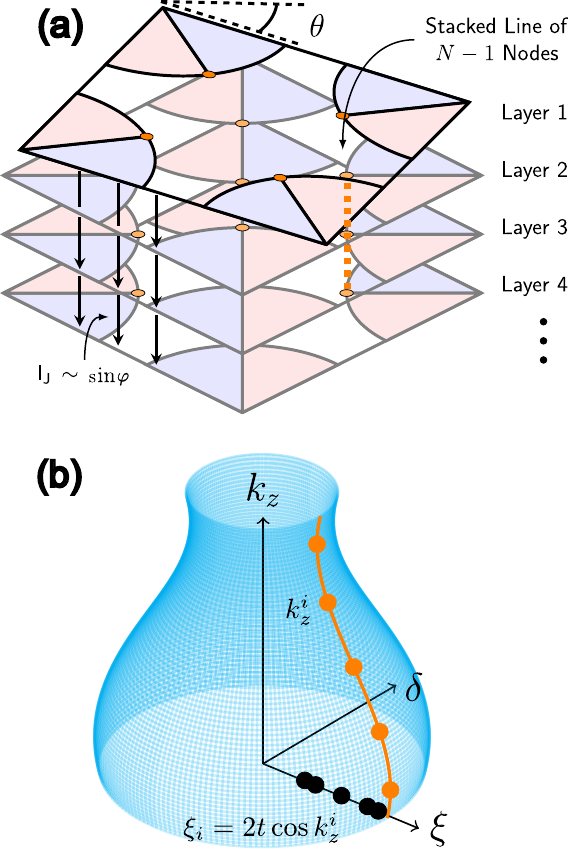}
    \caption{(a) Illustration of a $N$ layered flake of $d_{x^2+y^2}$ superconductors with a supercurrent $I_J$.
    (b) Emergence of Dirac nodes in $N$-layer flake without supercurrent from 3D ($N\to\infty$) Fermi surface in the space of $\xi$ (Eq.~\eqref{eq:xi}), $\delta$ (Eq.~\eqref{eq:delta}), and $k_z$. 
    A line node appears along $k_z$ in 3D highlighted by the orange curve. For a finite $N$, $k_z$ becomes quantized (see Eq.~\eqref{eq:h0eig}) illustrated by the orange points along the curve. Because of the quantization, $N$ Dirac nodes are introduced at corresponding values $\xi_i = 2t\cos k_z^i$ illustrated by the black points along $\xi$.
    }
    \label{fig:kz_nodes}
\end{figure}

In this section we consider the case of low twist angles $\theta\ll\theta_{MA}$ ($\delta_0\ll t$), where twist can be treated perturbatively. We find that in this case, while the topological gap is reduced with thickness, the coupling between all layers in the stack leads to collective effects beyond those expected of $N$ topological superconducting layers. We break the Hamiltonian of Eq.~\eqref{eqn:Hcont_full} down into three parts
\begin{equation}
        H({\bf k},\varphi) = H_0({\bf k}) + H_{\mathrm{cur}}(k_\perp,\varphi)
        +H_{\mathrm{twist}}(\varphi),
\end{equation}
and each term is expressed as:
\begin{equation}
    \begin{gathered}
        H_0({\bf k})
        =
        \sum_{n=1}^{N}  \Phi^\dagger_{n} (\xi \tau_3 +\delta \tau_1) \Phi_{n}   - t \sum_{n=1}^{N-1} \Phi^\dagger_{n}  \tau_3  \Phi_{n+1} + h.c. ,
    \\
    H_{\mathrm{cur}}(k_\perp,\varphi) = \sum_{n=1}^{N} \Phi^\dagger_{n}\delta [(\cos(\varphi n)-1)\tau_1+ \sin(\varphi n) \tau_2]\Phi_{n},
        \\
    H_{\mathrm{twist}}(\varphi) =  
    \Phi^\dagger_{1}\delta_0 [(\cos\varphi-1)\tau_1+ \sin\varphi \tau_2]\Phi_{1}.   
    \end{gathered}
    \label{eq:Hcont_part}
\end{equation}
$H_0({\bf k})$ describes the system without twist or current, while the remaining part is split into $H_{cur}(k_\perp,\varphi)$ which describes the effect of current through the flake and $H_{twist}(\varphi)$ for the current and twist of the $N=1$ layer. The eigenstates of $H_0({\bf k})$ (system in the absence of supercurrent and twist) are superpositions of plane waves along $z$, with open boundary conditions at $n=1$ and $n=N$. The latter are equivalent to adding two additional layers $n=0,N+1$ and requiring the eigenstates for the system to vanish at $n=0,N+1$. The resulting eigenstates are:
\begin{eqnarray}
   &    \Psi^\dagger_{k_i^z}   = \frac{1}{\sqrt{N+1}}\sum\limits_{n=1}^{N} \sin (k_i^z n) \psi^\dagger_{\bt{k},n}, 
    \label{eq:h0eig}
\end{eqnarray}
where $\psi^\dagger_{\bt{k},n}$ is a spinor composed of $ \psi^\dagger_{\bt{k},n} = \Phi^\dagger_{\bt{k},n} - \Phi^\dagger_{\bt{k},-n}$ and $k_z^i = \frac{i \pi}{N+1}$, $i =1,2,\dots, N $. In this basis, the Hamiltonian $H_0(\bt{k})$ takes the form:
\begin{equation}
    H_0(\bt{k}) = \sum_{i=1}^N 
    \Psi^\dagger_{k_z^i}
 \Big( \left( \xi - 2t\cos k_z^i \right)  \tau_3 
+\delta\tau_1 \Big) \Psi_{k_z^i},
    \label{eq:zero_en}
\end{equation}
with energies given by 
\begin{equation}
    E_{0}^\pm(\bt{k}) = \pm \sqrt{\left( \xi - 2t\cos k_z^i\right)^2+\delta^2}.
\end{equation}
Fig. \ref{fig:kz_nodes} (a) represents the model as seen in layer space and once transformed into $k_z$. In the $N \rightarrow \infty$ limit, the zero-energy manifold takes the form of a nodal line along the $z$-axis:  $\xi = 2t\cos k_z$, $\delta=0$, where $k_z\in[0,\pi]$. On the other hand, for finite number of layers the boundary conditions restrict $k_z$ to $N$ discrete values on this nodal line resulting in a discrete set of zero-energy (Dirac) points at $\xi_i = 2t\cos k_z^i$, $\delta=0$ (see Fig. \ref{fig:kz_nodes} (b)).

\subsubsection{Low current: $d+id$ proximity effect}
We now analyze Eq.~\eqref{eq:Hcont_part} in the limit $\varphi\ll 1/N$, focusing on the Dirac nodes identified above. In this limit,
$H_{twist}(\varphi)\approx \Phi^\dagger_{1}\delta_0  \sin\varphi\tau_2 \Phi_{1}$. Moreover, $H_{cur}(k_\perp,\varphi)$ vanishes at the node ($\delta=0$), while away from the nodes, it is much smaller than $\delta$ in this limit, and can thus be neglected when studying the states at energies near zero.

This case also corresponds to another experimentally relevant setup. Consider that the current is applied only between the top two layers (being fed in from the flake's side, as in Ref. \cite{Zhao2021}). Assuming that the in-plane critical current is much larger than the c-axis one, the in-plane phase gradient can be neglected resulting in the same problem, but without restriction on $\varphi$~\footnote{In this case, $H_{cur}(k_\perp,\varphi)$ vanishes identically while the $\tau_1$ term in $H_{twist}(\varphi)$ can be absorbed in a small shift of the Dirac point positions $\delta=0\to \delta=\delta_i\sim \frac{1}{N}$ without the requirement $\varphi \ll \frac{1}{N}$.}. Keeping this case in mind, we will keep $\sin\varphi$ in $H_{twist}(\varphi)$.

The problem has now reduced to a stack of untwisted d-wave superconductors coupled by tunneling with a small $id$ pairing component present in the top layer. The Hamiltonian in the small twist angle limit can be expressed as 
\begin{equation}
    \begin{gathered}
        H({\bf k}) = H_0({\bf k}) + H_{\mathrm{twist}}(\varphi), \quad \varphi \ll \frac{1}{N}
    \\
    H_{\mathrm{twist}}(\varphi) =  
    \Phi^\dagger_{1}\delta_0 \sin\varphi \tau_2 \Phi_{1}.   
    \end{gathered}
    \label{eq:hamforpert}
\end{equation}
Projecting $H_{twist}(\varphi)$ onto the eigenstates $H_0({\bf k})$ (Eq.~\eqref{eq:h0eig}),  we find terms acting within a single $k_z^i$ subspace that couple the two zero-energy states of the Dirac cones as well as those coupling different $k_z^i$. We neglect the latter term by applying $\delta_0 \sin\varphi \ll t$~
\footnote{Generally, in second-order perturbation theory there will be $O(N)$ terms of the order $\frac{1}{N} \frac{[\delta_0 \sin(\varphi)]^2}{2t (\cos k_z^i-\cos k_z^j)}$. Only $O(1)$ of them have the denominator of order $t/N$, while for the rest the denominator is of order $t$. So for $|\delta_0 \sin(\varphi)| \ll t$ the sum of all the terms will have $O(1)$ terms of order $\frac{[\delta_0 \sin(\varphi)]^2}{2t}$ and $O(N)$ terms of order $\frac{1}{N} \frac{[\delta_0 \sin(\varphi)]^2}{2t}$, allowing to neglect it for $\delta_0 \sin(\varphi)\ll t$.}
, which can be fulfilled for small enough twist angle and/or current. As a result, we get the effective Hamiltonian:
\begin{equation}
\begin{gathered}
    H_{\mathrm{eff}}(\bt{k},\varphi) = \sum\limits_{k_z^i}  \Psi^\dagger_{k_z^i}\Big( (\xi-\xi_i) \tau_3 + \delta \tau_1 + m(k_z^i,\varphi) \tau_2 \Big)   \Psi_{k_z^i} ,
    \\
    m(k_z^i,\varphi) = -\frac{2 \delta_0}{N+1}\sin \varphi   \sin^2 (k_z^i).
    \end{gathered}
    \label{eq:zero_E_eff_H}
\end{equation}
We therefore find that the presence of the $id$ component in the top layer "proximitizes" the whole flake, opening gaps at all Dirac nodes each with a mass given in Eq.~\eqref{eq:zero_E_eff_H}.

Importantly, the chirality of the nodes is the same, which for a two band continuum model corresponds to a Chern number of every points to be $\pm 1/2$~\cite{Bernevig2013}, so that they add up to $\pm\frac{N}{2}$. Furthermore, this model considers only the Dirac nodes arising in the vicinity of one original Dirac node of the 2D superconductor, but the total Chern number will depend on the sum of all nodes in the Brillouin zone. In fact, symmetry-related nodes in the Brillouin zone will also have the same Chern number \cite{Volkov2022}, resulting in 
\begin{equation}
   C_{\mathrm{tot},\varphi\ll 1/N} = N_{\mathrm{sym}} \frac{N}{2}, 
   \label{eq:chern_sym_total}
\end{equation}
where $N_{\mathrm{sym}}$ is to total number of symmetry-related  nodes in the Brillouin zone. For instance, $N_{\mathrm{sym}}=4$ for the case of d-wave (4 nodes related by $C_4$ symmetry). 
This prediction is consistent with $N$ layers being "proximitized" by the topological superconductivity emerging at the interface. Experimentally, this leads to an increase of the thermal Hall response of the system at low temperature discussed in detail in Sec. \ref{sec:exp}.




Another important characteristic of a chiral topological superconductor is the value of the gap opening at a Dirac node. Eq.~\eqref{eq:zero_E_eff_H} demonstrates that the gap values are different at different nodes, so that the system is described by a distribution of topological gaps, rather than a single one. For $1,i \ll N $, the topological gap scales as $m(k_z^i,\varphi)\propto \mathcal{O}(N^{-3})$. On the other hand, the typical gap scales as $\langle m(k_z^i,\varphi)\rangle_{\mathrm{typ}}\sim \mathcal{O}(N^{-1})$, which dominates the average gap value for $1\ll N$ given by $ \langle m(k_z^i,\varphi) \rangle =\frac{\delta_0 \sin \varphi}{N+1}$. Therefore, while the proximity-induced topological gap is reduced with the flake thickness, the reduction is only power-law like due to the gapless character of the unperturbed system. Combined with an increase in Chern number, this leads to robust experimental signatures of topology in flakes, discussed in Sec. \ref{sec:exp}.

\subsubsection{Finite Supercurrent: Topological Transitions in 2.5 D }
\label{sec:25dtop}
We now move onto the case with finite current (arbitrary $0\leq\varphi\leq\pi/2$ in Eq.~\eqref{eq:Hcont_part}). In the preceding section, we've demonstrated that the low-energy spectrum at $\varphi\ll1/N$ takes the form of gapped Dirac nodes at $\delta=0$. Let us now consider what happens in this case on increasing $\varphi$. In particular, $H_{cur}(k_\perp,\varphi)$ in Eq.~\eqref{eq:Hcont_part} can no longer be neglected. Projecting $H_{cur}(k_\perp,\varphi)+H_{twist}(\varphi)$ onto the basis of Eq.~\eqref{eq:zero_en} we obtain:
\begin{equation}
\begin{gathered}
    H_{\mathrm{eff}}
    = \sum_{k_z^i}   \Psi^\dagger_{k_z^i} 
    \begin{pmatrix}
        \xi- 2t\cos k_z^i & D(\varphi)\\
        D^*(\varphi) & - \xi+ 2t\cos k_z^i
    \end{pmatrix}   \Psi_{k_z^i}, 
    \\
    D(\varphi,k_z^i) = \delta D_\delta(\varphi,k_z^i) +\delta_0 D_m(\varphi,k_z^i) ,
  \\
D_m(\varphi,k_z^i) = \frac{2}{N+1} e^{i\varphi} \sin^2 k_z^i ,
   \\
   D_\delta(\varphi,k_z^i)  = \frac{ \sin(\varphi\frac{N+1}{2})e^{i\varphi\frac{N+1}{2}} }{N+1} \\
  \times \frac{\sin^2 k_z^i \cot \frac{\varphi}{2}}{\sin(k_z^i - \frac{\varphi}{2})\sin(k_z^i + \frac{\varphi}{2})}.
    \end{gathered}
    \label{eq:hamproj_SC}
\end{equation}
At low $\varphi\ll 1/N$ we recover the previous result of $N$ gapped Dirac nodes. However, on increasing $\varphi$ we find that this structure breaks down in two cases:

\underline{$(i)\; D_\delta(\varphi) = 0:$}  in this case the dispersion of a Dirac node flattens along the $k_\perp(\delta)$ direction, such that the dispersion is no longer Dirac-like. This behavior occurs near quantized values of phase $\varphi_\delta^j $,
\begin{equation}
    \varphi_\delta^j = \frac{2 \pi j}{N+1}, 
    \label{eq:transition_1}
\end{equation}
where $ j = 1 , 2, \dots, N$. Not all Dirac nodes are affected, rather, only those in the range $i  \in \{i =j , j+1,\dots, N+1-j \}$. In other words, the corresponding nodes in the range positioned along $\xi_i$ will have this broadening dispersion while those for $i < j$ and $i > N+1-j$ will not. As $\varphi$ is induced by the Josephson current, its maximal allowed value per Josephson relation is $\pi/2$, limiting the values of $j$ to be below $\frac{N+1}{4}$. Furthermore, for $N\leq 3$, 
$\varphi_\delta^1 \ge \pi/2$, implying that this situation can not be realized in bi- or tri-layer flakes.

\begin{figure}[h!]
    \centering
   \includegraphics[width=0.4\textwidth]{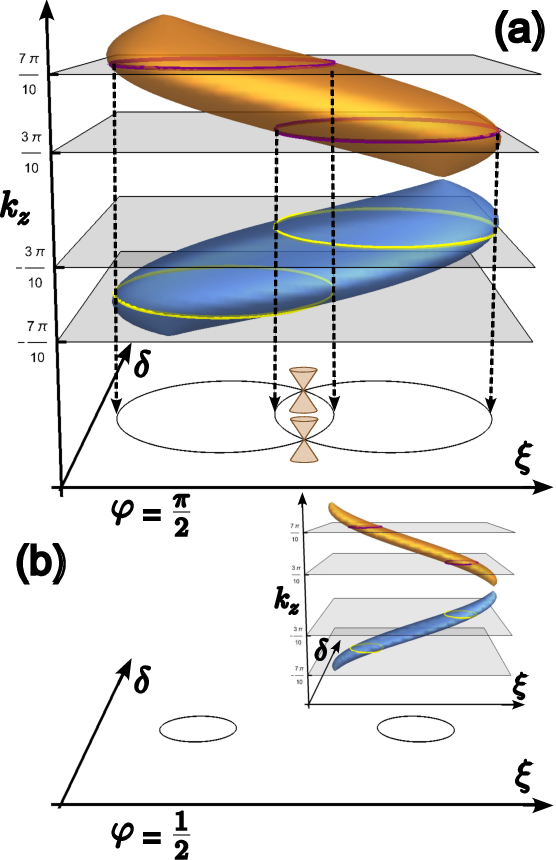}
    \caption{Generation of Dirac nodes in a superconducting flake under c-axis supercurrent $I = I_c \sin \varphi$. (a) Top: zero-energy contour of a bulk superconductor under $\varphi=\pi/2$ in the space of $\xi$, $\delta$, and $k_z$ as defined in Eqs.~\eqref{eq:xi} and \eqref{eq:delta}. Planes show four values of $k_z$ allowed by open boundary conditions in $N=9$ layer flake. Bottom: Projection of zero-energy contours for $k_z$ marked in the top panel. 2D Dirac  points form at the points of their intersection. (b) For smaller value of $\varphi$, the contours with the same $k_z$ no longer intersect, and no Dirac points are formed.}
    \label{fig:current_nodes}
\end{figure}

Importantly, once $\varphi$ equals $\varphi_\delta^j$, Dirac nodes at $\xi_i$ and $k_\perp = 0$ are recovered, and persist between $\varphi_\delta^j$ and $\varphi_\delta^{j+1}$.
As shown below, along with the recovered Dirac nodes, an additional pair appear in the spectrum along $k_\perp(\delta)$ which have split from each node positioned at $k_\perp = 0$ and $\xi_i$. An interesting consequence is the impact on the chirality of the new pair and recovered Dirac nodes. Once the transition occurs, the recovered Dirac nodes change sign while the new pair preserve the chirality of the Dirac node prior to the splitting (see Fig. \ref{fig:current_nodes}). In this way, the overall chirality of the three nodes after the transition remains the same as before the transition.

\underline{$(ii)\; D_\delta(\varphi) = r D_m(\varphi),\;  r \in \mathbb{R} : $} in this case, $D_m(\varphi)$ acts as a shift of $\delta$, and can be absorbed into $D_\delta(\varphi)$ which closes the gap. For values of $\varphi$ across this transition, the chirality of all the $\delta = 0$ nodes flip. This occurs when ${\rm Im} [D_\delta(\varphi) D_m^*(\varphi)]=0$ equivalent to:
\begin{equation}
    \varphi^j = \frac{2\pi j }{N-1}, \; j = 1,2,\dots
    \label{eq:transition_2}
\end{equation}
Once $\varphi >  \frac{2\pi j }{N-1}$, the chirality of the nodes flip relative to when $\varphi < \frac{2\pi l }{N-1}$, signifying that the chirality of all nodes for $\delta = 0$ reverse sign. 
Note that since $\varphi \leq \frac{\pi}{2}$ due to the Josephson relation, the first time this transition can occur is for $N  = 6 $. From hereon, every additional four layers adds another possible transition (i.e. $N=10$ can have two transitions, $N=14$ can have three, etc.).

As has been suggested above, the system at finite current bias may host additional Dirac points not close to $\delta=0$. To study this, we consider the zero-energy eigenstates of $H_0(\bt{k})+H_{cur}(k_\perp,\varphi)$ at $\delta\neq0$ adding the effect of $H_{twist}(\varphi)$ perturbatively. By applying a transformation $c_{n}^\dagger \rightarrow e^{-i\varphi n/2} c_{n}^\dagger$, $H_0(\bt{k})+H_{cur}(k_\perp,\varphi)$ can be brought to the form:
\begin{eqnarray}
H_0(\bt{k})& +& H_{\mathrm{cur}}(k_\perp,\varphi) \to 
 \sum_{n=1}^{N}  \Phi^\dagger_{n} (\xi \tau_3 +\delta \tau_1) \Phi_{n} \nonumber  \\
& & - t  \sum_{n=1}^{N-1} \Phi^\dagger_{n}  e^{i\varphi n/2}\tau_3  \Phi_{n+1} + h.c. 
        \label{eq:ham_gauge_trans}
\end{eqnarray}
Away from the boundaries, the eigenstates of Eq.~\eqref{eq:ham_gauge_trans} are plane waves $|k_z,\varphi,\uparrow/\downarrow\rangle = \frac{1}{\sqrt{N}}\sum_{n=1}^N e^{i k_z n}|n,\uparrow/\downarrow\rangle$ satisfying
\begin{equation}
\begin{gathered}
 H_{\delta\neq 0}(\bt{k}) |k_z,\varphi\rangle=
   E_{{\bf k},k_z,\varphi} |k_z,\varphi \rangle
   ,\\
 H_{\delta\neq 0}(\bt{k})  = 
 \left(\xi - 2 t \cos k_z \cos \frac{\varphi}{2} \right) \tau_3   + \delta \tau_1 \\
 - 2 t \sin k_z \sin \frac{\varphi}{2} \tau_0.
\end{gathered}
\label{eq:ham3dcur}
\end{equation}
For Eq.~\eqref{eq:ham3dcur}, this Hamiltonian has corresponding energy eigenvalues of

\begin{eqnarray}
    E^\pm_{{\bf k},k_z,\varphi} & = & -2 t \sin k_z \sin \left(\frac{\phi }{2}\right) \nonumber \\
    & & \pm\sqrt{\delta ^2+(\xi- 2 t \cos k_z \cos \frac{\phi}{2})^2}.
    \label{eq:ham3dcur_eigen}   
\end{eqnarray}
For each value of $\xi<2 t$ we find $4$ zero-energy eigenstates when we satisfy the condition:

\begin{equation}
 \cos k^{\pm}_z = \frac{\xi \cos(\frac{\phi}{2}) \pm \sqrt{(4t^2-\xi^2)\sin^2(\frac{\phi}{2}) - \delta^2  } }{2t},
 \label{eq:cosk}
\end{equation}
where $k^{\pm}_z$ are the corresponding momenta to form the zero-energy eigenstates.

For an infinite layered system, Eq.~\eqref{eq:cosk} is the equation for zero-energy states which form a closed 2D surface (see Fig. \ref{fig:current_nodes}). We now introduce open boundary conditions as in the case leading to Eq.~\eqref{eq:h0eig}. As shown, the boundary conditions at $n=1,N$ result in a quantization condition $k_z^i = \frac{i \pi}{N+1}$. Moreover, since the boundary conditions results in $4$ equations, a solution only exists if there are $4$ states with the same energy for a given $\xi,\delta$. In the case of zero-energy states, the $k_z^i$ quantization selects discrete ``rings" in $\xi, \delta$ space (see Fig. \ref{fig:current_nodes}). Additionally, the Hamiltonian from Eq.~\eqref{eq:ham3dcur} possesses a particle-hole symmetry which guarantees that each ``ring" is doubled. However, as noted above, satisfying the boundary conditions requires $4$ independent solutions. This arises when the projections of two rings to the $\xi,\delta$ plane cross (see Fig. \ref{fig:current_nodes} (a)). This corresponds to quantizing the momenta  in Eq.~\eqref{eq:cosk} as $k^{n,+}_z = k_z^i + k_{\delta} = \frac{n\pi}{N+1}$
and 
$k^{m,-}_z = k_z^i - k_{\delta} = \frac{m \pi}{N+1}$
with arbitrary integer $ 1 \leq n,m \leq N$. These conditions require quantization of $k_\delta$ as
 $k^j_\delta \in \{ \frac{j \pi}{N+1} | 1 \leq j \leq N$ \}. As explored in the Appendix \ref{appendix:quant}, the quantization condition for $k^{\pm}_z$ and $k_{\delta}$ implies that the new Dirac nodes exist when:

\begin{equation}
\min[\sin^2 k_z^i ,\sin^2 k^j_{\delta}]<\sin^2[\varphi/2]<\max[\sin^2 k_z^i ,\sin^2 k^j_{\delta}],
\label{eq:cond}
\end{equation}
where $k^j_\delta < k_z^i$ to avoid $k^{m,-}_z< 0 $ which correspond to an inequivalent value of $k^j_\delta$. 
For $j=1$, an additional Dirac node exists for all $i \neq 1, N$ while $\frac{2\pi}{N+1}<\varphi<2k_z^i$. Note that the number of such crossings depends on $\varphi$, as some crossings may disappear and new ones appear (see Fig. \ref{fig:current_nodes} (b)). Since the particle-hole symmetry above holds for finite flakes, there will always be a pair of zero-energy solutions.

\begin{figure*}[t]
         \includegraphics[width=.95\textwidth]{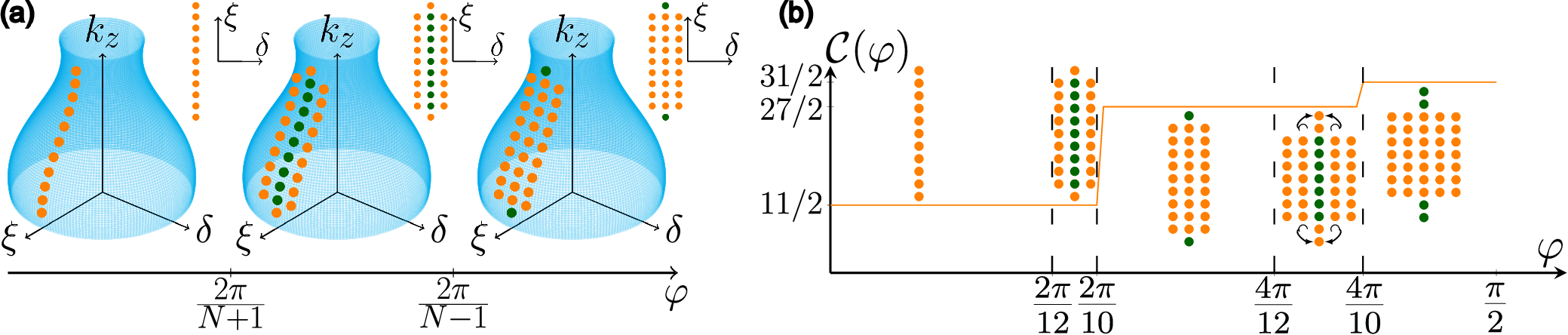}    
\caption{Schematic of current-driven evolution of the quasiparticle dispersion for $N=11$. (a) Evolution of the gapped Dirac node number, position and chirality (marked by color) with increasing $\varphi$, related to interlayer supercurrent (cf. Fig. \ref{fig:kz_nodes}). Insets show the projection of the nodes in the 2D space of $\delta$ and $\xi$. 
b) illustrates the mechanism of current-driven topological transitions in the $\xi-\delta$ plane. Increasing the current nodes along $\xi_i$ split and form pairs of nodes along $\delta$. At $\varphi = \frac{2\pi}{10},\frac{4\pi}{10}$ the $\delta=0$ nodes change chirality, resulting in topological phase transition which increases the Chern number. The dashed circles denote nodes which merged with a node along $k_\perp=0$ shown by the arrows at $\varphi=\frac{4\pi}{12}$.
}
\label{fig:node_evol_cart}
\end{figure*}

We can now demonstrate that each zero-energy point in $\xi,\delta$ space corresponds to a gapped Dirac point with the same Chern number value. Using an orthonormal pair of eigenstates formed from the zero energy eigenstates, one can project onto the momenta local to the nodes. To induce the topological phase, we perform this projection to the boundary mass of Eq.~\eqref{eq:Hcont_part} as well. By performing these projections, one can form a low energy effective Hamiltonian with the form
\begin{equation}
    H^{\delta \neq 0}_{\mathrm{eff}}(\bt{k}) = k_i \mathcal{A}_{ij} \sigma_j + m(\varphi,\delta_{0},k_z^i) \sigma_2,
    \label{eq:multidirac}
\end{equation}
where $k_i$ are the vector components of $(k_\parallel,k_\perp,k^i_z)$ 
, $\sigma_j$ is the zero energy basis, $m(\varphi)$ is the effective mass which also depends on $t$, and $\mathcal{A}$ is a $3\times3$ matrix defined in Appendix \ref{appendix:lowE_nonzero_delta} which generalizes the mixing of $\xi,\;\delta$ between basis components $\sigma_1,\;\sigma_3$. In the limit $\varphi \ll 1$, the twist and current induced Dirac mass term takes the form
\begin{equation}
    m(\varphi,k^j_\delta,k_z^i) \approx - \frac{2 \delta_0}{N+1} \frac{k^j_\delta}{\varphi} \sin^2 k_z^i.
    \label{eq:m_delta=0}
\end{equation}
The corresponding Chern number for each node is then $\mathcal{C} = \frac{\textrm{sign}(m(\varphi,k^j_\delta,k_z^i))}{2}$. 
Thus, we arrive at the conclusion that for finite $\varphi$ one can generate new Dirac nodes with the same chirality.

To illustrate the behavior of the Dirac nodes, we present the sequence of evolutions to the number and chirality of the Dirac nodes as we increase $\varphi$ from zero. For a specific example, we illustrate the sequence for $N=11$ layers in Fig. \ref{fig:node_evol_cart}. Once $\varphi>0$, all Dirac nodes acquire a mass and all have the same chirality. Increasing further, we reach the first condition  $\varphi = \frac{2\pi}{N+1}$ (i.e. case $(i)$ is satisfied), where we enter a region for  $\frac{2\pi}{N+1}<\varphi<\frac{2\pi}{N-1}$ where new Dirac nodes split off for all but the extreme nodes (i.e. those corresponding to $\pm k_z^1 $). From the previous analysis, the original net Chern number determined for $0<\varphi<\frac{2\pi}{N+1}$ follows from the sign of the mass term in Eq.~\eqref{eq:zero_E_eff_H}, and is preserved even when the nodes along $\xi_i$ split. To do so, nodes along $\xi_i$ which split flip their chirality to $-\textrm{sign}(\delta_0)$, while the nodes at $\pm k_\delta$ preserve chirality of $\textrm{sign}(\delta_0)$. Therefore, the total chirality of the three nodes is the same as the node prior to splitting.

The next condition is when $\varphi=\frac{2 \pi}{N-1}$ (case $(ii)$ is satisfied). For this case, a gap closing occurs which results in a topological phase transition. Increasing $\varphi$ further such that $\frac{2\pi}{N-1} < \varphi < \frac{4\pi}{N+1}$, all $\xi_i$ nodes for $k_\perp = 0$ reverse chirality due to the gap closing at $\varphi = \frac{2\pi}{N-1} $, but not the pair of nodes formed from splitting. As a result, there's a net change in Chern number before and after the closing of $\Delta \mathcal{C}_{net} = N-4$. To summarize the discussion so far, we've completed a sequence for $j=1$ in Eq.~\eqref{eq:transition_1} and Eq.~\eqref{eq:transition_2} where nodes split while remaining in the same topological phase, and then increase the Chern number for increasing $\varphi$.

If we continue to increase $\varphi$, we enter the second sequence for $j=2$ where the cycle repeats. Note, however, that upon entering a new sequence, the nodes corresponding to $\pm k_z^i$ for $i=j,N+1-j$ vanish by merging with a node along the $k_\perp=0$ axis as illustrated in Fig. \ref{fig:node_evol_cart} (b) for $N=11$. Otherwise, continuing to $j=2$, for $\frac{4\pi}{N+1}< \varphi < \frac{4\pi}{N-1}$, a new pair of node splitting occurs so long as $\frac{4\pi}{N+1} < \varphi < 2k_z^i $ (i.e. all nodes of $k_z^i$ split except the four at $k_z^i = \pm \frac{\pi}{N+1}, \pm \frac{2\pi}{N+1}$) with the same effects on their chirality. Once we increase $\varphi$ to the next topological phase transition, as before, the Chern number changes but by $\Delta \mathcal{C}_{net} = N - 8$. 

Overall, completing each sequence results in topological transitions which compound to increase the Chern number. 
 We can represent the Chern number as a function of $\varphi$ as:
\begin{equation}
\mathcal{C}(\varphi) = \frac{N}{2} + \sum\limits_{n=1}^{\lfloor \frac{N-1}{2\pi}\varphi \rfloor} 
\Big( N-4n \Big)  ,\;\; N \geq 6 
\end{equation}
where $\lfloor \dots \rfloor$ is the floor function used to count the number of transitions that occurred for a given value of $\varphi$ (i.e. the index $j$ for which  $\varphi^j=\frac{2 j \pi}{N-1}$ and $\lfloor \frac{N-1}{2\pi}\varphi \rfloor = j$). For example, taking $\varphi=\pi/2$ and $N\gg 1$ we get $\mathcal{C}(\varphi)\approx \frac{3}{32}N^2$. Therefore, with respect to the number of layers, the Chern number scales as $\mathcal{O}(N^2)$.

\subsection{Large twist angles}
\label{subsec:large_twist}

We now consider the opposite case $\theta \gg \theta_{MA}$. We first use the continuum model (Eq.~\eqref{eqn:Hcont_full}), that is valid in this regime provided that $\theta_{MA}\ll 1$. In this case, the tunneling between top two layers can be treated perturbatively, as the respective Dirac points are far apart for $\delta_0\gg t$. The Hamiltonian of the top layer is given by
\begin{eqnarray}
    H_{sl}(\bt{k}) &= &\phi^\dagger_0 \Big( \xi \tau_3  + (\delta + \delta_0) \cos \varphi \tau_1  \nonumber \\
        & & - (\delta + \delta_0) \sin \varphi \tau_2 \Big) \phi_0,
\end{eqnarray}
and for the stack, we will utilize the Hamiltonian from Eq.~\eqref{eq:zero_en}. We now introduce a tunneling term between layer $0$ and layer $1$
\begin{equation}
    H_t(\bt{k}) = -t\phi^\dagger_0  \tau_3 \phi_1 + h.c.
\end{equation}
Applying second order perturbation theory to the zero energy eigenstates produces mass terms 
\begin{eqnarray}
   H_{m,\mathrm{pert}}(k_z^i,\varphi) =  -\phi^\dagger_0 \big( \sum_{k_z^i} m(k_z^i) \tau_1 \big) \phi_0  \nonumber \\
   + \phi^\dagger_{k_z^i} m(k_z^i) \big( \cos\varphi \tau_1 - \sin\varphi \tau_2 \big)\phi_{k_z^i},
\end{eqnarray}
for $i = 1,2,\dots, N$ and the mass term is given as
\begin{equation}
    m(k_z^i) = \frac{2}{N+1}\frac{t^2\sin ^2 k_z^i}{\sqrt{4t^2\cos k_z^i +\delta_0^2 } }.
\end{equation}
Treating $\delta_0 \gg t$, then  $m(k_z^i) \approx \frac{2}{N+1} \frac{t^2}{\delta_0} \sin^2 k_z^i $ and $\sum_{k_z^i} m(k_z^i) \approx \frac{t^2}{\delta_0}$. The end result is that the Dirac point of the top layer acquires a gap with magnitude independent of $N$, while the Dirac gaps in the stack follow the low twist behavior with gaps $\mathcal{O}(\frac{1}{N})$ to $\mathcal{O}(\frac{1}{N^3})$ at large flake thickness.

\begin{figure}[t]
    \includegraphics[width=.45\textwidth]{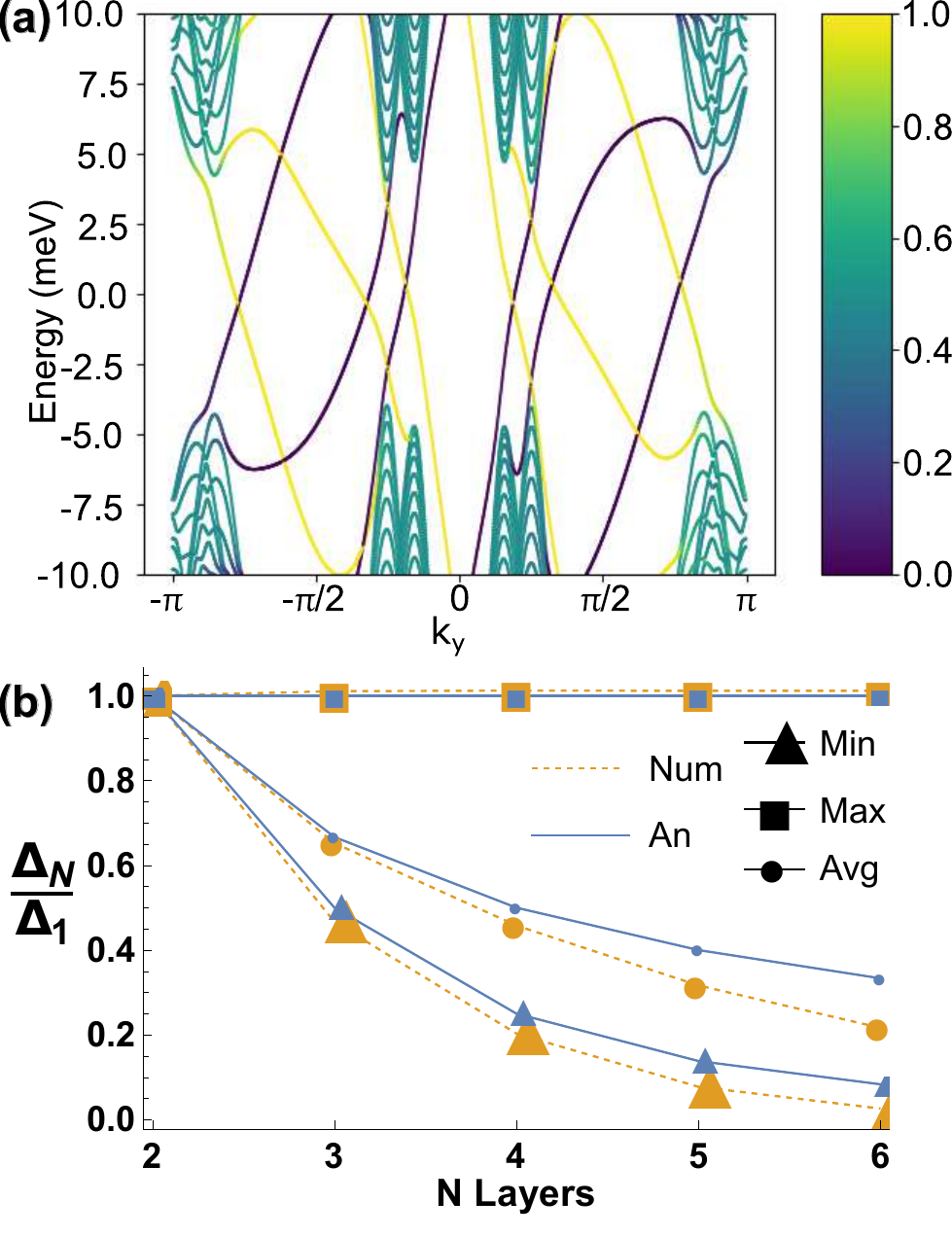}
    \caption{a) Quasiparticle dispersion for a 120 unit cell strip of the $N=3$ flake for $\theta=\theta_{1,3}\approx36.9^\circ$ and $\varphi_c = \frac{\pi}{2}$ is plotted. The color scale represents the expectation value of position along the open direction. The edge modes signify the presence of topological gap corresponding to a Chern number of 6.(b) Topological gaps as a function of $N$ for $\theta=\theta_{1,2}\approx 53.1^\circ$, $t = 5$ meV and high twist continuum model. The gap values are normalized by their value for $N=2$.
    The maximal gap is almost $N$-independent, reflecting that the twisted layer is largely decoupled from the remaining stack at large twist.}
    \label{fig:moire_dispersion_LDOS}
\end{figure}

In this limit, we can compare the results with the numerical solution of the tight-binding model, Sec. \ref{subsec:lattice}. We compute the spectrum numerically on a strip of the moir\'{e} superlattices with open boundary conditions along one axis. In Fig. \ref{fig:moire_dispersion_LDOS} (a), we show the resulting eigenmode spectrum for the $N=3$ and $\theta_{1,3}$ structure. The chiral edge modes appear inside a bulk gap, consistent with chiral topological superconductivity. Moreover, the number of modes (6 per edge) is exactly as expected in the case of a ``proximitized'' flake discussed above. Explicitly computing the Chern number from the bulk following the definition in Appendix \ref{appendix:Chern} matches the number of modes/edge exactly and corresponds to a Chern number of 6. In Fig. \ref{fig:moire_dispersion_LDOS} (b), we present the dependence of the minimal, maximal and average topological gaps on $N$. Remarkably, the analytical and numerical results agree well. Note that the analytical model does not take into account the realistic Fermi surface geometry and moir\'{e} reconstruction of the bands, while the lattice model does, indicating a non-trivial robustness of topological superconductivity in twisted flakes.

\section{Experimental Predictions}
\label{sec:exp}

\begin{figure}
    \includegraphics[width=.42\textwidth]{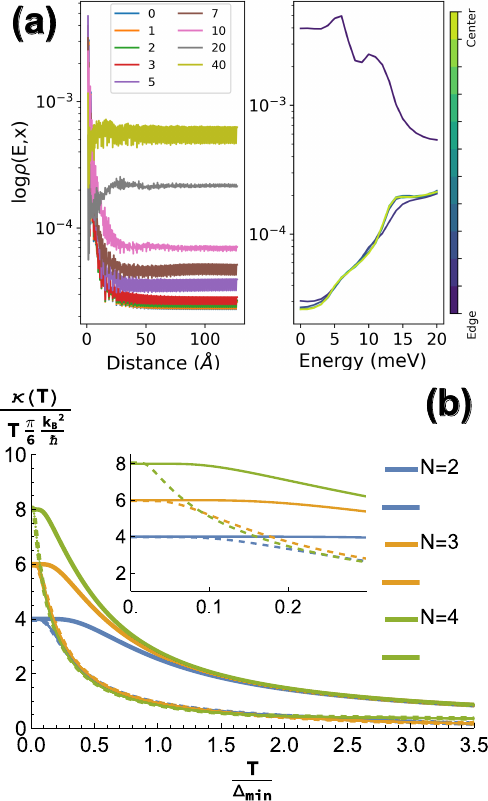}   
    \caption{(a) Local density of states on the top layer of a 2+1 layer 120 unit cell strip with angle $\theta_{1,3}$ and $t = 80$ meV as a function of position along the open axis at fixed energies. The right panel displays the LDOS versus energy for fixed positions between the left and center of the system. LDOS shows a pronounced increase in the bulk above a threshold energy, while below it, LDOS is concentrated at the edges indicating presence of a topological gap. 
    (b) Normalized thermal Hall coefficient as a function of temperature normalized by the $N=2$ topological gap. The conductance was computed at a cutoff of $|E_{cutoff}| = 5$ meV.}
    \label{fig:TH_num_and_an}
\end{figure}

We now consider the implications of the topological superconductivity in flakes for two types of experiments: scanning tunneling microscopy (STM) and thermal Hall effect (THE) \cite{Volkov2022}. 
%
STM gives access to the local density of states of the system as a function of energy. Choosing energies within the topological gap, the contribution of edge states can be separated. The presence of a topological phase can then be identified by a zero local density of states (LDOS) in the bulk of the systems with nonzero LDOS localized at the boundaries. As the topological gap of the top layer can be almost $N$-independent (Fig. \ref{fig:moire_dispersion_LDOS} (b)), this opens the path to detecting topological superconductivity in thin/thick flake structures.


THE, on the other hand, provides a measurement of the system's total Chern number at low temperatures \cite{Senthil1999,Kallin_2016,Volkov2022}. At finite temperatures, THE can appear even in non-topological twisted superconductors \cite{Hu2023THE}, however, the magnitude of the effect is smaller than for the topological ones. Most importantly, it can be used to detect the topological phase transitions, predicted in Sec. \ref{sec:25dtop}.

\subsection{Scanning Tunneling Microscopy}

Here we analyze the local density of states (LDOS) of the top layer defined as:
\begin{eqnarray}
    \rho(E,x) &=& 
    \sum\limits_{\substack{\bm{k} \in BZ \\ i \in L }}\sum\limits_{n} \Big( | \langle u_{i}(x_i) | n \rangle|^2 \delta(E-E_n) \nonumber \\
    && + | \langle v_{i}(x_i) | n \rangle|^2 \delta(E+E_n) \Big),
\end{eqnarray}
where $i$ enumerate the sites which are positioned at the same coordinate $x_i$ in a given layer $L$. Furthermore, $| n \rangle$ are the energy eigenstates with associated energy $E_n$. The eigenstates $| n \rangle$ are expressed in terms of Nambu space with $|u_{i}(x_i)\rangle$ and $|v_{i}(x_i)\rangle$ representing the particle-like and hole-like components, respectively, for each site.

In Fig. \ref{fig:TH_num_and_an} (a) we present the LDOS in the 3 layered system of the microscopic model (Sec. \ref{subsec:lattice}). A noticeable LDOS is observed only near the system boundary while the interior of the system appears empty for energies below a threshold. For additional clarification, we also show the LDOS for a fixed position as the energy bias varies. Here, the center of the system has zero LDOS until the bias is larger than the gap size, where bulk bands start to contribute. On the other hand, the edges show considerable LDOS at zero bias that actually diminishes on increasing the bias. 

We note that the LDOS is only strictly zero below the minimal topological gap of the flake. To illustrate this point, we compute the projected density of states (PDOS), as outlined in Appendix \ref{appendix:LDOS}, on the top layer of the Hamiltonian in the large twist limit outlined in Sec. \ref{subsec:large_twist} using the eigenstates corrected to first order. The PDOS, $\rho_0(E)$, for the top layer is given as
\begin{equation}
    \rho_0(E) = \frac{E}{2\pi v_F v_\Delta},\quad E \geq 2\frac{t^2}{\delta_0}\sin\varphi,
    \label{eq:PDOS_0}
\end{equation}
which remains gapped even in the $N \rightarrow \infty$ limit. However, the $N$ layer stack also provides a contribution of 
\begin{equation}
    \rho_{N}(E) = \frac{E}{\pi v_Fv_\Delta} \frac{1}{N+1} \frac{t^2}{\delta_0^2} \mathlarger{\sum}\limits_{\substack{ k_z^i\\E \geq m(k_z^i)\sin\varphi } } \sin^2 k_z^i,
    \label{eq:PDOS_N}
\end{equation} 
which becomes gapless for large $N$. However, for large twist angles, this contribution is suppressed, resulting in a clear signature at the top layer's gap energy, as confirmed by our numerical results (Fig. \ref{fig:TH_num_and_an}).

\subsection{Thermal Hall effect}

We now consider the thermal Hall (TH) conductivity $\kappa_{xy}(T)$, that is directly related to the Chern number $\mathcal{C}$ of the system at low temperatures via $\kappa_{xy}(T)\approx-\frac{\pi}{6}CT\frac{k_B^2}{\hbar}$\cite{Senthil1999,Kallin_2016,Volkov2022}.
To compute $\kappa_{xy}(T)$ we follow the approach of~\cite{Cvetkovic2015,Volkov2022}
(see Appendix \ref{appendix:TH} for details). In Fig. \ref{fig:TH_num_and_an} (b) , we present $\kappa_{xy}(T)$ normalized by $\frac{\pi}{6}T\frac{k_B^2}{\hbar}$ computed for the microscopic model of $\theta_{1,2} $ structure and for the low-angle continuum model for up to $N=4$. For the continuum model, we used the Dirac approximation from Eq.~\eqref{eq:multidirac} (see also Appendix \ref{appendix:lowE_nonzero_delta}). At low temperatures, all curves reach a plateau with value equal to the Chern number and width related to the energy gap of the system. Comparing results from both models in Fig. \ref{fig:TH_num_and_an}, they plateau to the same values, but the reduced energy gap in the lattice model diminishes the width of the plateau relative to the continuum. Interestingly, away from low temperature, $\kappa_{xy}(T)$ appears independent of the number of layers. This result also follows from the continuum model (Sec. \ref{sec:cont}) in the $1 \ll N $ and $m(\varphi,k_\delta,k_z^i) \ll 1$ limit where $m(\varphi,k_\delta,k_z^i)$ is defined in Eq.~\eqref{eq:m_delta=0}, we find
\begin{equation}
    \kappa_{xy}(T) \approx  \delta_0\frac{3}{2 \pi^2} \frac{\log 2}{k_B T} , \; N \rightarrow \infty.
    \label{eq:TH_largeN}
\end{equation}
This result arises because, on the one hand, the average topological gap decreases $\propto 1/N$, but at the same time the Chern number grows $\propto N$. At temperatures much larger then the gap, $\kappa_{xy}\propto \mathcal{C}_{net} \langle\Delta\rangle/T$ where $ \langle\Delta\rangle$ is the average gap size, and is thus independent of layer number. This opens the possibility of using finite-thickness flakes to study the appearance of topological SC at the twisted interface.

\subsubsection{Topological Phase Transitions}
We now demonstrate, that the current-driven topological transitions (Sec. \ref{sec:25dtop}) can be studied with the THE.
Each transition will correspond to an abrupt increase in Chern number, reflecting in a larger $\kappa_{xy}(T)$ value at low temperatures. To evaluate the TH coefficient, we use the continuum model, Sec. \ref{sec:25dtop} and use the Dirac approximation for low energies. In particular, we take use the gaps in the Dirac approximation for nodes along $\delta=0$ from Eq.~\eqref{eq:hamproj_SC} and for $\delta\neq0$ from Eq.~\eqref{eq:m_delta=0} applied to the TH coefficient defined in Appendix~\ref{appendix:TH}. We can then sum together the TH coefficient from each node which determines $\kappa_{xy,tot}(T)$. 

\begin{figure}[h]

    \begin{minipage}{.35\textwidth}
         \includegraphics[width=\textwidth]{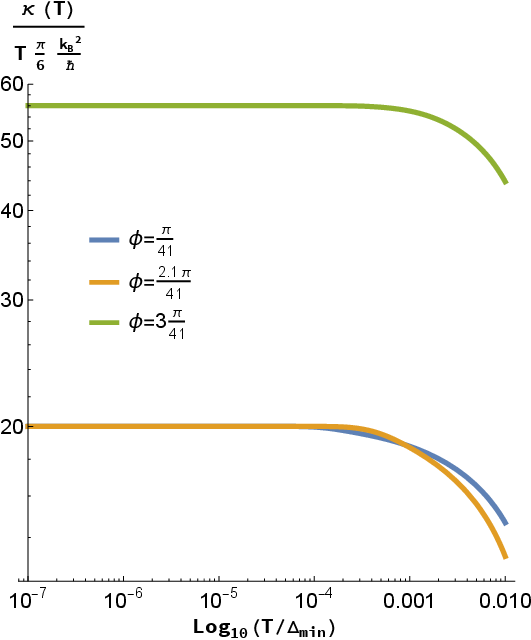}        
    \end{minipage}
\caption{ 
TH coefficient for the $N=40$ layered structures for various values of $\varphi$. Plotted is the $\kappa_{xy}(T)$ normalized by its low temperature limit $\kappa^0_{xy}(T\to0)$ versus the temperature normalized by the gap of the $N=2$ layered system computed numerically using the continuum model. The abrupt increase in plateau value with $\varphi$ tracks the current-driven topological phase transition, Sec. \ref{sec:25dtop}.
}
    \label{fig:node_evolution_TH_an_transition}
\end{figure}

The results in the inset of Fig. \eqref{fig:node_evolution_TH_an_transition} show the effect on $\kappa_{xy,tot}(T)$ at low temperature (and therefore the Chern number) in the case $\varphi,\frac{\delta_0}{2t} \ll 1$ and  for $N= 40$ layers in the lowest sequence (i.e. $j=1$ in the conditions from Eq.~\eqref{eq:transition_1} and Eq.~\eqref{eq:transition_2}.).  When $\varphi < \frac{2\pi}{N+1}$, we have a topological gap and a Chern number of $20$, consistent with the contribution of a single gapped Dirac node for 40 layers. Once $\varphi$ first passes $\frac{2\pi}{N+1}$, the Dirac nodes for all $k_z^{i}$ except for $k_z^{\pm 1} \neq \pm \frac{\pi}{N+1}$ split, producing a pair of nodes. However, as shown in Sec. \ref{sec:25dtop}, the gap does not close during this process and the Chern number remains the same.
One observes also that the temperature width of the plateau grows, reflecting the overall increase in the gap value similar to what is expected for weak currents.


As the phase is then increased past $\frac{2\pi}{N-1}$ (see Fig. \ref{fig:node_evolution_TH_an_transition}, green curve), the value of the plateau suddenly changes. The magnitude of change is $\Delta \mathcal{C}_{net} = N-4 = 36$, as predicted in Sec. \ref{sec:25dtop} due to the gap closing for Dirac nodes along $\delta=0$. 
The plateau range concomitantly increases, reflecting a larger value of the topological gap. 
Compared to the results in Eq.~\eqref{eq:TH_largeN}, the fact that the Chern number $\mathcal{O}(N^2)$ once $\varphi = \frac{\pi}{2}$, then $\kappa_{xy,tot}(T) \approx  \frac{N}{T}$ allowing for a distinction between the measurements of $ \kappa_{xy,tot}(T)$ for each topological phase transition.

\section{Conclusion} 
\label{sec:conclusion}

In this work we have demonstrated that finite-thickness flakes of nodal superconductors can be used to realize chiral topological superconductivity using current. We identified that the signatures of the topological phase, such as thermal Hall effect or local density of states reconstruction can be almost independent of the flake thickness (Sec. \ref{sec:exp}) for a monolayer twisted on top of a thick flake. This platform is an easier alternative to fabricate compared to twisted bilayers, encouraging experimental outlook. Moreover, we have shown that increasing current (Sec. \ref{sec:25dtop}) leads to a sequence of topological transitions, leading to an overall increase in Chern number up to $\sim \mathcal{O}(N^2)$. Such states clearly transcend the expectations for $N$ independent layers of 2D topological superconductors and suggests a new phase that can be realized only in finite-thickness flakes. As such, this provides an example of a ``2.5"-dimensional effect---one that cannot be observed in a bulk system (due to decreasing topological gap) but that scales with system's thickness in a nonlinear way. Combining these observations with effects of magnetic field will open a way to generate even more elaborate 2.5D topological superconductors and manipulate their properties~\cite{Volkov2022}, and promote future prospects of other 2.5D moir\'{e} and heterostructure systems.

\acknowledgements{We thank John Bonini, Marcel Franz, Philip Kim, Abhay Pasupathy, and Justin Wilson for useful discussions. This work is partially supported by NSF Career Grant No.~DMR- 1941569 and the Alfred P.~Sloan Foundation through a Sloan Research Fellowship (K.L., J.H.P.). Part of this work was performed in part at the Aspen Center for Physics, which is supported by the National Science Foundation Grant No.~PHY-2210452 (J.H.P.,P.A.V.) as well as the Kavli Institute of Theoretical Physics that is supported in part by the National Science Foundation under Grants No.~NSF PHY-1748958 and PHY-2309135 (J.H.P., P.A.V.).}
\appendix

\section{Quantization Condition for BdG Dirac Nodes}
\label{appendix:quant}
From Eq.~\eqref{eq:cosk} and treating $k_\pm = k_z^i \pm k_\delta $, we can form the relations
\begin{equation}
\begin{gathered}
 \cos k_+ + \cos k_- =  2 \cos k_z^i \cos k_\delta = \frac{\xi \cos\frac{\varphi}{2}}{t},
 \\
 \cos k_+ - \cos k_- = - 2 \sin k_z^i \sin k_\delta= \sqrt{\sin^2 k_z^i \sin^2\frac{\varphi}{2} - \delta^2}.
 \end{gathered}
\end{equation}
Squaring the expressions,
\begin{eqnarray}
& \cos^2 k_z^i \cos^2 k_\delta = \frac{\xi ^2\cos^2(\frac{\varphi}{2})}{4t^2},
\label{eq:xi_1}
 \\
& \sin^2 k_z^i \sin^2 k_\delta= \sin^2 k_z^i \sin^2\frac{\varphi}{2} - \delta^2.
\label{eq:xi_2}
\end{eqnarray}
Substituting $\frac{\xi^2}{4t}$ from Eq.~\eqref{eq:xi_1} into Eq.~\eqref{eq:xi_2}

\begin{equation}
\begin{gathered}
 \sin^2 k_z^i \sin^2 k_\delta= \sin^2\frac{\varphi}{2} - \tan^2\frac{\varphi}{2}\cos^2 k_z^i \cos^2 k_\delta- \delta^2,
 \\
  \sin^2 k_z^i \sin^2 k_\delta + \tan^2\frac{\varphi}{2}\cos^2 k_z^i \cos^2 k_\delta 
  -\sin^2\frac{\varphi}{2}
  = - \delta^2<0.
 \end{gathered}
\end{equation}
The latter inequality results in a quadratic inequality for $\sin^2\frac{\varphi}{2}$ which we solve to form the inequality

\begin{equation}
\min( \sin^2 k_z^i ,\sin^2 k_\delta ) < \sin^2\frac{\varphi}{2} < \max( \sin^2 k_z^i ,\sin^2 k_\delta ).
\label{eq:cond_app}
\end{equation}
Note that $k_\delta< k_z^i$ is required, otherwise $k_+ < 0 \equiv \pi - k_+$, which corresponds to a different value of $k_\delta$. From the quantization conditions, we see for $k_\delta = \frac{n\pi}{N+1}$, we get a Dirac point at $k_0$ so long as $\frac{2n\pi}{N+1}< \varphi< 2 k_z^i$. If $\varphi>2 k_z^i$, then the Dirac point vanishes and merges with a node along $\delta=0$.

\section{Low Energy Effective Hamiltonian}
\label{appendix:lowE_nonzero_delta}
Performing projections of the zero energy eigenvectors of the Hamiltonian in Eq.~\eqref{eq:ham3dcur} leads us to an effective Hamiltonian of the form

\begin{eqnarray}
    & H_{eff}^{\delta\neq 0}(\bt{k}) = \Big( a_\xi ( \xi - \xi_0) + b_\delta (\delta - \delta_0) \Big) \sigma_3 + \Big( b_\xi( \xi - \xi_0) \nonumber \\
    & + a_\delta (\delta - \delta_0) \Big) \sigma_1 - i m(k_z^i,k^\pm_z)\sigma_2 ,
    \label{eq:App_H_eff}
\end{eqnarray}
where $a_\xi$, $a_\delta$, $b_\xi$, and $b_\delta$ are constant factors expressed as

\begin{eqnarray}
 a_\xi & = & \frac{\alpha^2_+}{2\beta^2_+}+ \frac{\alpha^2_-}{2\beta^2_-} ,\\
b_\delta & = & \frac{\alpha_+}{2\beta^2_+}  +  \frac{\alpha_-}{2\beta^2_-} ,\\
b_\xi & = & - \frac{\alpha_+}{2\beta^2_+} -\frac{\alpha_-}{2\beta^2_-} , \\
a_\delta & = & -1,
\end{eqnarray}
where
\begin{eqnarray}
    \alpha_\pm & = & \frac{2t}{\delta_0}(\cos k_z^i - \cos k^\pm_z \cos\frac{\phi}{2} ) ,\\
    \beta_\pm & = & \frac{2t}{\delta_0}\sqrt{ (\cos k_z^i - \cos k^\pm_z \cos\frac{\phi}{2} )^2 +\frac{\delta_0^2}{4^2}  } ,
\end{eqnarray}
$m(k_z^i,k^\pm_z)$ is an effective mass, and $\xi_0$ and $\delta_0$ are the dispersions corresponding to the zero energy solution. To simplify the Hamiltonian, we can write the terms corresponding to the BdG Dirac node from the mass term produced by the twist as

\begin{equation}
   H_{eff}^{\delta\neq 0}(\bt{k}) =  k_i \mathcal{A}_{ij} \sigma_j + i m \sigma_2 ,
\end{equation}
where $k_i$ is a momentum component corresponding to $(k_\parallel-k_{\parallel,0}, k_\perp-k_{\perp,0},k_z)$ and $\mathcal{A}_{ij}$ is a $3\times3$ matrix where $i,j = 1,2,3$. Corresponding to the effective Hamiltonian in Eq.~\eqref{eq:App_H_eff}, the matrix $\mathcal{A}$ is expressed as

\begin{equation}
    \mathcal{A} = \begin{pmatrix}
    v_F b_\xi  & 0 &  v_F a_\xi \\
    v_\Delta a_\delta  & 0 &  v_\Delta  b_\delta \\
    0  & 0 &  0
    \end{pmatrix} .
\end{equation}

\section{Layer Projected Density of States}
\label{appendix:LDOS}

Here we outline the PDOS when restricted to a layer in the continuum picture. As in the lattice LDOS, contributions will be split between particle- and hole-like components but only restricted to a given layer, 

\begin{eqnarray}
    \rho_l(E) &=& \int \frac{dk_\parallel dk_\perp}{4\pi^2}  \sum\limits_{n} \Big( | \langle u_{l}(x) | n \rangle|^2 \delta(E-E_n') \nonumber \\
    && + | \langle v_{l}(x) | n \rangle|^2 \delta(E+E_n') \Big),
\end{eqnarray}
where $(k_\parallel,k_\perp)$ compose the coordinate system, $| n \rangle$ indexes the eigenstate and corresponding energy eigenvalue $E_n$, and $l$ is the layer index we are performing the projection onto. If performed on a lattice, the integral will be replaced by a sum over a k-mesh in the Brillouin zone normalized by the number of points in the mesh.

\section{Chern Number}
\label{appendix:Chern}

The Chern number is numerically evaluated by integrating the Berry curvature
\begin{equation*}
\mathcal{C} = \frac{1}{2\pi} \int_{BZ} \Omega( \bt{k} ) d^2 k.
\end{equation*}
In this formula, the Berry curvature is defined in a gauge-invariant fashion as 

\begin{equation*}
\Omega( \bt{k} ) = \sum_{m\neq n} \frac{\langle n | \nabla_{\bt{k}} H(\bt{k}) | m \rangle \times \langle m | \nabla_{\bt{k}} H(\bt{k}) | n \rangle }{(E_m-E_n)^2},
\end{equation*}
where $|n \rangle$ and $| m \rangle$ are energy eigenstates with corresponding energy eigenvalues $E_n$ and $E_m$, respectively.

%
\section{Thermal Hall}
\label{appendix:TH}
%

The normalized thermal Hall (TH) conductivity is computed as

\begin{equation}
    \kappa_{xy}(T) =  \frac{1}{\hbar T} \int_{-\infty}^\infty d\zeta \zeta^2  \Big( -\frac{\partial f(\zeta)}{\partial \zeta} \Big) \sigma_H(\zeta),
    \label{eq:TH_formula}
\end{equation}
where $f(\zeta)$ is the Fermi-Dirac function and $\sigma_H(\zeta)$ is the Hall conductance defined for various energy filling. For the microscopic model, $\sigma_H(\zeta)$ is expressed as

\begin{eqnarray}
\sigma_H(\zeta) & & =  -i \int_{BZ} \frac{ d^2 k}{4\pi^2} \\
& & \times \sum_{m\neq n}\sum_{E < \zeta}  \frac{\langle n | \nabla_{\bt{k}} H(\bt{k}) | m \rangle \times \langle m | \nabla_{\bt{k}} H(\bt{k}) | n \rangle }{(E_m-E_n)^2}. \nonumber
\end{eqnarray}
The normalized TH conductivity can be simplified to the form

\begin{equation}
    \kappa_{xy}(T) =  \frac{6}{4 \pi T^3} \int_{-\infty}^\infty d\zeta \sigma_H(\zeta) \frac{\zeta^2}{\cosh^2(\frac{\zeta}{2T})},
\end{equation}
which will normalize as $T \rightarrow 0$ to the Chern number $|C|$. Here we compute the TH conductivity within a quadrant of the Brillouin zone (relying on $C_4$ symmetry), resulting in a low temperature value of $|C|/4$.\\

\subsection{Continuum Two Band Model}
The TH coefficient for the continuum model can be computed using a two band description of a massive Dirac Hamiltonian. A general massive Dirac Hamiltonian can be expressed as~\cite{Bernevig2013}

\begin{equation}
    H(\bf{k}) = \sum_i d_i(\bt{k})\sigma_i, 
\end{equation}
where the vector components of ${d}_i(\bt{k})$ for the continuum model can be expressed as

\begin{equation}
    \qquad d_1(\bt{k}) = \delta, \quad d_2(\bt{k}) = m, \quad d_3(\bt{k}) = \xi,
\end{equation}
with $d = \sqrt{\sum_i {d}_i(\bt{k})^2} $ and $m$ is an arbitrary mass parameter. Here, we begin by considering a generic mass term $m$. In this form, the Berry curvature can be written as

\begin{equation}
    \Omega_{\zeta \delta}(\bt{k}) = \frac{1}{2d^3} d_2(\bt{k}) \partial_\xi d_3(\bt{k}) \partial_\delta d_1(\bt{k}).
\end{equation}
Substituting for the vector components, the conductance for a given energy $\zeta$ can be expressed as

\begin{equation}
    \hat{\sigma}_H(\zeta) =  \frac{1}{4\pi^2}\frac{1}{v_Fv_\Delta} \int d k_\parallel d k_\perp \frac{m}{4d^3},\;\; |\zeta| < d.
\end{equation}
The integral can now be evaluated piece-wise to yield,

\begin{equation}
    \hat{\sigma}_H(\zeta)= \frac{1}{2\pi}
    \begin{cases}
    \textrm{sign}(m), \quad & |\zeta| < m \\
    \frac{m}{|\zeta|},  & |\zeta| \geq m
    \end{cases}.
\end{equation}
Using the conductance, we can compute the TH coefficient in Eq.~\eqref{eq:TH_formula} which is normalized by its low temperature limit to yield $\kappa_H^0(\beta)$ expressed as

\begin{eqnarray}
  &   \kappa_{xy}^0(\beta)   =
  \frac{6}{\pi^2}\beta^2\Big(  - \frac{m}{\beta}\log \left(2 \cosh \left(\frac{\beta m}{2}\right)\right) \nonumber \\
  &+\frac{1}{\beta^2}\Big( 2 \text{Li}_2(-e^{-\beta m} )+\frac{\pi ^2}{6} \Big) +\frac{m^2}{2}  \Big),
\end{eqnarray}
where $Li_s(z)$ is the polylogarithm with $s$ the order of its series expansion. So far, $m$ has been kept as an arbitrary parameter, but we wish to evaluate the TH coefficient for all nodes. Let's consider the case of nodes along $\delta=0$ at values of $k^i_z$ and mass corresponding to a given node will be represented as $m(k^i_z)$. Treated in a continuum limit of layers where $1 \ll N $ and $m(k^i_z) <1$, we have an integral over $k_z \in (0,\pi)$,

\begin{eqnarray}
    &\kappa_{xy,tot}^0(\beta) 
    = \frac{1}{N}\int\limits_0^N dn \kappa_H^0(\beta,m(k^i_z) ) \nonumber \\
    & \approx \frac{1}{\pi} \int\limits_0^{\pi} dk_z \kappa_H^0(\beta,m(k_z) ).
\end{eqnarray}
Evaluating this integral gives us Eq.~\eqref{eq:TH_largeN} when we substitute for an explicit expression for $m(k_z)$ (e.g. $m(k_z) = \frac{\delta_0}{N} \sin^2 k_z )$.

\bibliography{manuscript_v5}

\end{document}